\documentstyle [11pt,epsf]{article}  
\title{Neutrino and Parity Violation in Weak Interaction  \\[1.5 cm] }
\author{ \it{ Kiyoung Kim} \\ 
       {\it Department of Physics, University of Utah, SLC, UT 84112\/}
        \\[1.5 cm] }   
\date{ }
\begin{document}
\maketitle
\begin{abstract}
\noindent
It is shown that neutrino oscillation and Majorana type neutrino are not 
compatible with Special Theory of Relativity. Instead of the Majorana type 
neutrino, traditional neutrino ($m_{\nu} = 0$) is considered with additional 
assumptions that $\nu_{_R}$ and  $\bar\nu_{_L}$ neutrinos exist in nature 
and naturally thus parity is conserved.  Besides, neutrino itself is 
considered as {\it longitudinal vacuum-string oscillations\/}.    
To explain parity conservation  $W^{\pm}$ bosons are suggested as momentary 
interacting states between  $e^{\pm}$ and virtual-positive(negative)-charge 
strings, respectively. 
With these postulations, an alternative explanation is also suggested for 
solar neutrino problem. 
\end{abstract}
\vspace{10 pt} 
\section{Introduction}
\paragraph{}
As the most amazing and even paradoxical one in elementary particles, neutrino 
has been known to us for a long time since 1931, when Wolfgang Pauli proposed 
a new particle -- very penetrating and neutral -- in $\beta$-dacay.  
Nevertheless, we don't know much about neutrino itself.  That is mainly 
because of experimental difficulty to detect neutrino itself and neutrino 
events too.  Up to now, we are not quite sure if neutrino has a rest mass or 
not, and also we cannot explain {\it solar neutrino deficit\/}\cite{Winter}
\cite{Kamio}\cite{SNE}\cite{Sun}   
contrary to the expectation from Standard Solar Model\cite{Sun}.  
Furthermore, Nature doesn't seem to be fair in {\it Weak Interaction\/}  
in which neutrino is involved.  That has been known as {\it Parity Violation\/}.  
\paragraph{}  
As one of trial solutions to explain the {\it solar neutrino deficit\/},  
neutrino {\it mass\/} or {\it flavor\/} oscillation has been suggested, and 
many experiments have been done and doing until now.  However, in the theory 
of neutrino flavor oscillation, it is supposed that neutrino, for instance, 
{$\nu_e$, $\nu_{\mu}$, or $\nu_{\tau}$}, is not in a fundamental state in 
Quantum mechanical point of view, but in a mixed state with {\it mass 
eigenstates\/}, such as {$m_1$, $m_2$, and $m_3$} ($m_1\not=m_2\not=m_3$).  
This supposition implies that neutrino has an intrinsic structure and, thus,  
one of flavors spontaneously transforms to another kind of flavors, like the 
conversion of Meson $K^o$ into its antiparticle $\bar K^o$.  
If neutrino flavor mixing is true, we have another dilemma in {\it lepton 
number\/} conservation, which has been quite acceptable in phenomenology.   
In section(1.2), we are going to compare neutrino flavor mixing  with the 
$K^o$ and $\bar K^o$ mixing.  Moreover, the theory itself will be reviewed 
if it is physically sound or not.    
\paragraph{}
In Weak Interaction, we know that {\it parity\/} is violated.  
When C. S. Wu {\it et al\/} showed parity violation in their  experiment\cite
{Wu}, it was astonishing because we had believed in a fairness in natural 
phenomena.   Up to now, we don't know the reason for {\it parity violation\/}, 
and also we cannot explain why {\it parity violation\/} is detected only in 
Weak Interaction. 
Should we accept this fact as nature itself?  Otherwise, there should be a 
reasonable reason.   
\paragraph{}
As mentioned above, the experimentation is difficult because neutrinos  
have extremely small cross-sections(typical cross-section : $10^{-44}$ $m^2$ 
(1GeV)\cite{Perkins}); furthermore, neutrinos participate in only Weak 
Interaction in which the interaction range and the relative strength are $\ll 1 
fm$($\sim 10^{-18} m$) and $10^{-15}$, respectively.\cite{Greiner}  Meanwhile, 
if the interaction range and the relative strength are compared with 
Electromagnetic Interaction case(range : $\infty$, relative strength : 
$10^{-2}$)\cite{Greiner}, it is not easy [for us] to understand why 
Electromagnetic Interaction - intermediated by {\it photon\/} -- is suppressed 
in the Weak Interaction range despite that its interaction range is $\infty$ 
and the relative strength is $10^{-2}$ which is $10^{13}$ times bigger than 
Weak Interaction one.\cite{Hans}  
\subsection{Fermion and Boson}
\paragraph{}
Through the Complex Space model\cite{Kim-1}, it was possible to understand how 
Special Theory of Relativity and Quantum Mechanics can be connected, 
what is the physical entity of wave function -- ontological point of view 
-- in Quantum Mechanics, and  how Schr{\"o}dinger equation is related to 
the Complex Space model\cite{Kim-2}.     
In that, Dirac's {\it hole theory\/}\cite{Dirac} was re-interpreted 
as following : In the Complex Space, vacuum particles -- vacuum electrons 
in the model -- are not packed completely; they can transfer energy(for 
instance, electromagnetic energy) through a wave motion such as 
{\it vacuum-string-oscillation\/}.  In which the electromagnetic energy is 
propagated along the imaginary axis, and it is realized to real space through 
U(1) symmetry.  
\paragraph{}
In Quantum Physics, we know that the {\it phase factor\/} of wave function is 
arbitrary if we consider only  probability density($\psi\cdot \psi^{*}$); 
yet, it need to be considered in physical interactions.  
Especially, its peculiarity was shown by Berry, M.V. in  considering slowly 
evolving Quantum phase in the interaction of fermion(for example, electron) 
with magnetic field, and it has been known as  
{\it Berry phase\/}.\cite{Berry}\cite{Thomas} 
\par
\medskip    
Now, let us think about what the electron's spin is and, in general, what the 
fermion(electron, proton, neutron, etc) is.  First of all, we know that 
the electron's spin can be represented as a rotation in {\it two dimensional 
complex space\/}, that is SU(2); and the spin vector is realized through an 
interaction in which the interaction Hamiltonian is $\alpha (\vec\sigma \cdot 
\vec r)$($\sigma$:pauli matrix, $\alpha$:constant).\cite{Berry}  Furthermore, 
we can physically surmise that the peculiarity of Quantum phase occurs when 
the spin vector in SU(2) is realized through the interaction with magnetic 
field because of the difference between SU(2) and SO(3) -- 
two to one correspondence in their parameter spaces even though both are 
similar groups.  Here, we can be sure, at least, that electron's spin vector(
axial vector) resides in the Complex Space. 
\begin{figure}[h] 
\begin{center}
\leavevmode
\hbox{%
\epsfxsize=4.9in
\epsffile{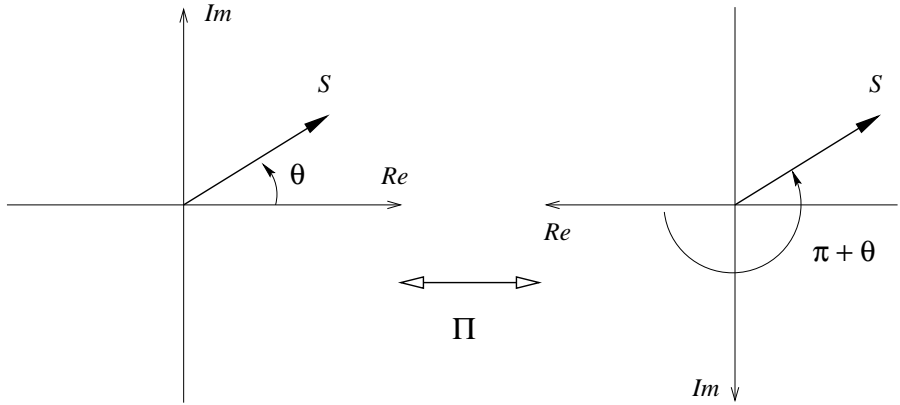}} 
\end{center}
\caption{phase change through space inversion} 
\label{Fermion}
\end{figure}
According to the interpretation of Quantum formalism in the Complex Space
\cite{Kim-2}, Quantum phase corresponds to the phase itself in the Complex 
Space.  Although the Quantum phase in a stationary system is arbitrary -- U(1) 
-- and doesn't affect physical interpretation, in a dynamical system the phase 
is no more free to choose. As a specific and simple example of Berry phase, 
let us imagine following : One electron(spin $\hbar/2$) is at origin under 
the influence of magnetic field($\vec H$) which is on $x$-$y$  plane with 
pointing to the origin. Now, let the field pointing direction rotate slowly 
around $z$ axis; then, the system is in SO(2) $\subset$ SO(3) in real space 
and U(1) $\subset$ SU(2) in the Complex Space because of the correspondence 
between SU(2) and SO(3); U(1) and SO(2) in their transformations.  Here, U(1) 
is different from the one we traditionally use since our concern is U(1) in 
the internal Complex Space.  Now let us say, the electron's spin points to 
the positive $x$ axis at the beginning.  And let the field direction rotate  
with $\phi$ degrees which corresponds to $\phi /2$ rotating simultaneously 
in the internal Complex Space.  
After a round trip of the system($\phi = 2\pi$) in real space, we can easily 
recognize that the spin points to the other direction -- the negative $x$ 
axis.      
\paragraph{}
From Pauli exclusion principle we know that any two identical fermions cannot 
be at the same Quantum state, but bosons can share the same state.  
That means, the wave function representing two identical fermion system  
is anti-symmetry, but the wave function of two identical bosons is 
symmetry.  To demonstrate this fact, let us say, $\Psi(1,2)$ represents  
the Quantum state for two identical fermions.  Then, $\Psi(2,1) = - \Pi
\makebox[0.4 mm]{}\Psi(1,2) =  - {\rm P}\makebox[0.4 mm]{}\Psi(1,2)$, 
where $\Pi$ and ${\rm P}$ are respectively {\it exchange\/} and {\it parity\/} 
operator  and it was used the fact that exchange operator($\Pi$) is 
identical to parity operator(${\rm P}$) for this two identical fermion system. 
In Fig.(\ref{Fermion}) total spin vector(axial) in two identical fermion 
system is in internal Complex Space.  After space inversion or parity 
operation the Quantum phase is changed from $\theta$ to $\pi + \theta$ with 
showing how the anti-symmetric wave function(two-fermion system) and the 
parity operation are connected.         
\paragraph{}
With this fact  we can distinguish {\it fermion\/} and {\it boson\/}. Fermion 
has a spin -- intrinsic angular moment(half integer $\times \hbar$) --  
in Complex Space; boson can also have a spin, but the spin vector
(axial) is in real space.     
\subsection{Neutrino  Oscillation}  
\paragraph{} 
One of long standing questions related elusive particle, neutrino, is neutrino 
oscillation.  To understand the theory of neutrino oscillation, 
firstly we have to assume that each flavor of neutrinos($\nu_{e}, \nu_{\mu}, 
\nu_{\tau}$)  has a rest mass. Of course, $m_{\nu_{e}} \not = 
m_{\nu_{\mu}} \not = m_{\nu_{\tau}}$ to be meaningful in the theory.  
Furthermore we have to suppose that there are fundamental Quantum states, 
in which the individual mass eigenstate doesn't appear in phenomenon, but 
linear combinations of these states emerge to $\nu_{e}, \nu_{\mu}$ or 
$\nu_{\tau}$.   
\paragraph{}
{\bf In the respect of Quantum formalism :\/} To describe {\it one\/} particle 
system, the wave function is a linear combination with all possible Quantum 
states -- orthogonal and complete set of bases.    
There, each state represents {\it the particle \/} itself  with a probability. 
Now, if the particle changes its identity, for instance, from $\nu_{e}$ to 
$\nu_{\mu}$, can we still use the same complete basis set to describe the new 
one?  Or should we change the basis set to new one also?  
According to the theory, we can use the same basis set to describe these 
phenomenologically different particles  because these particles are 
fundamentally identical but in different states to each other.  
However, $K^o$ and $\bar K^o$ mixing is different from neutrino flavor mixing 
case because they have internal structures and thus can be different intrinsic 
parities. In phenomenon, they appear to have different properties and 
transform spontaneously to each other with exchanging $\pi^{\pm}$(Pions).  
Now what about neutrinos? Do they have been known they have internal 
structures and different intrinsic parities?   
Although it has been known that neutrinos have imaginary intrinsic 
parities\cite{Boehm}, it doesn't make any sense in phenomenology.   
What about in the middle of oscillation?  For instance, 
``$50 \%$ $\nu_{e}$ and $50 \%$ $\nu_{\mu}$'' is possible?  Although we can 
use Quantum formalism to describe a physical system, the particle's identity  
in the physical system should be clear at any time.  Otherwise, 
there should be an additional explanation why the fundamental mass states 
are beyond physical phenomena.     
\paragraph{}
{\bf Energy and Momentum conservation:\/} Let us say, rest mass  $m_{\nu_{e}}$ 
-- electron neutrino -- is moving in an isolated system with velocity 
$\beta_e$ at $t = 0$, and after $\tau$ seconds($\tau$ is arbitrary)  
the mass is changed to rest mass $m_{\nu_{\mu}}$($m_{\nu_{\mu}} \not= 
m_{\nu_{e}}$) -- muon neutrino --  and the velocity to $\beta_{\mu}$.  
First of all,  energy conservation should be satisfied as 
\begin{equation}
m_{\nu_{e}} \gamma_e = m_{\nu_{\mu}} \gamma_{\mu} 
\label{E-conser}
\end{equation}
where $\gamma_e = 1/\sqrt{1-\beta_{e}^2}$, 
      $\gamma_{\mu} = 1/\sqrt{1-\beta_{\mu}^2}$, and  $ c \equiv 1$, 
Moreover their momentums also should be conserved as 
\begin{equation}
m_{\nu_{e}} \gamma_e \beta_{e} =  m_{\nu_{\mu}} \gamma_{\mu} \beta_{\mu} 
\label{P-conser}
\end{equation}
If the masses, $m_{\nu_{e}}$ and $m_{\nu_{\mu}}$ are different, both equations,
 Eqn.(\ref{E-conser}) and Eqn.(\ref{P-conser}), cannot be satisfied 
simultaneously.  
\paragraph{}
Even though our concerning is different in Special Theory of Relativity and 
Quantum Physics, both theories should be equally satisfied by a new theory 
because they are basic theories in physics and connected fundamentally.
\cite{Kim-1}         
\section{Dirac equation and Majorana neutrino}
\paragraph{}
To describe spin ${1\over 2} \hbar$ particle(fermion), two formalisms -- 
Dirac\cite{Itzykson_1}\cite{Lewis} and Majorana\cite{Majorana}\cite{Case}
\cite{Boehm} formalism -- have been known.  
Since both formalisms are not derived directly from Sch{\"o}dinger equation --
that is not deductive but inductive, it is natural [for us] to check if these 
two formalisms are physically enough meaningful or not.  
\paragraph{}
In Majorana case; neutrino and antineutrino are identical such as  
$\nu_{_L} \equiv \bar \nu_{_L}$ and  $\nu_{_R} \equiv \bar \nu_{_R}$
(Majorana abbreviation), 
and neutrino has a rest mass($m_{\nu} \not = 0$).  Although the first 
condition is tolerable in the respect of phenomenological facts that we 
couldn't have detected  $\bar \nu_{_L}$ and $\nu_{_R}$, the second one is not 
compatible with Special Theory of Relativity.  
\par
\medskip  
If $m_{\nu} \not = 0$ no matter how small it is\footnote{our main concern here 
is about neutrino.}, then we can find the neutrino rest frame through proper  
Lorentz transformation; furthermore, we can even flip the helicity 
from right-handed($\bar\nu_{_R}$) to left-handed($\nu_{_L}$) or {\it vice versa\/}. 
Because there are only two kinds of neutrinos, $\nu_{_L}$ and $\bar \nu_{_R}$,  
the flipped one is always corresponded to the other one.     
However, obviously we can distinguish $\nu_{_L}$ from $\bar \nu_{_R}$ in their 
properties.\cite{Cowan}\cite{LA}   
For instance, in interactions; $\nu_{\mu} + e  \longrightarrow \nu_{\mu} + e$ 
and $\bar \nu_{\mu} + e  \longrightarrow \bar \nu_{\mu} + e$, 
their total cross sections, $\sigma_{\nu e}$ and $\sigma_{\bar \nu e}$, are 
different.\cite{LA}  If Majorana neutrinos can be applicable in physics, they 
should be identical in their physical properties, that is, 
$\nu_{_L} \equiv \bar \nu_{_R}$.  Yet, the total cross section can be changed 
suddenly at a critical boost velocity in the transformation in spite of 
the fact that the cross section should be invariant.  That means,  a physical 
fact, which should be unique in all Lorentz frames, can be changed through the 
transformation.  Therefore, both conditions in Majorana formalism cannot be 
feasible in physical situation as long as Special Theory of Relativity is 
impeccable.  About the neutrino mass, there was already a similar argument in 
the respect of Group Theory in 1957\cite{Lee}.   Therefore, neutrinos satisfied 
by both conditions are not appropriate in physics.   
\par
\medskip 
If we abandon the first condition but still assume that neutrino mass is not zero, 
we have to treat neutrinos as like other spin ${1\over 2} \hbar$ fermions -- 
not Majorana type fermions.   
However, what if we give up the mass of neutrino but hold the first condition, 
$\nu_{_L} \equiv \bar \nu_{_L}$ and $\nu_{_R} \equiv \bar \nu_{_R}$?  Then, we have 
to ask again the old questions if $\nu_{_R}$ and $\bar\nu_{_L}$ exist or not; 
if these two neutrinos exist, why we cannot detect them; why parity is violated 
in Weak Interaction.  As a possible case, let us assume that neutrino has no mass
($m_{\nu} = 0$) and that $\nu_{_R}$ and $\bar\nu_{_L}$ exist but we couldn't have 
detected them yet.  If this supposition, which corresponds to Majorana 
neutrino with setting $m_{\nu}$ to zero\cite{Case}, is true, we can understand 
why parity violation happens in Weak Interaction; moreover,  we can find 
a clue to understand {\it solar neutrino deficit\/} problem.     
\par
\medskip 
In Dirac formalism; four spinors are closed by {\it parity\/} operation in 
general, and in neutrino case($m_{\nu} = 0$) the Dirac equation is decoupled 
to two component theory -- Weyl equations.  To compare spinor parts among 
neutrinos($\nu_{_R}$, $\nu_{_L}$, $\bar\nu_{_R}$, and $\bar\nu_{_L}$) 
in free space,  let us say :   
\begin{eqnarray}
\nu_{_R}(x)& = & \nu_{_R}(k) e^{-ikx} \makebox[1 cm]{;}  
     \nu_{_L}(x) \makebox[2 mm]{} = \makebox[2 mm]{} \nu_{_L}(k) e^{-ikx} 
     \nonumber        \\ 
\bar\nu_{_R}(x) & = &  \bar\nu_{_R}(k) e^{ikx} \makebox[2 mm]{}\makebox[1.0 cm]{;}
     \bar \nu_{_L}(x) \makebox[2 mm]{} = \makebox[2 mm]{} \bar\nu_{_L}(k) e^{ikx} 
     \label{X-K}  
\end{eqnarray}
for a given $k$. Here, $\nu_{_R}$ represents positive energy with right-handed 
neutrino; $\bar\nu_{_L}$, negative energy with left-handed antineutrino, 
and {\it etc\/}.    
Using {\it charge conjugation\/} operator $C$  which is consistent with 
the Dirac's hole interpretation,  we can see that there are only two independent 
solutions\cite{Itzykson-2}  as   
\begin{eqnarray}
\bar\nu_{_R}(k)  &=& C \gamma^o \nu^*_L(k)   \makebox[2 mm]{}  
                 = - \nu_{_R}(k)    \makebox[2 cm]{and} \nonumber  \\ 
\bar\nu_{_L}(k) & = & C \gamma^o \nu^*_R(k) \makebox[2 mm]{}
                 = - \nu_{_L}(k),  
\label{C-C}  
\end{eqnarray} 
in which $\gamma^o\makebox[1 mm]{}=\makebox[1 mm]{}
\pmatrix{ & I   & 0   \cr 
          & 0   & -I  \cr} $ and $C\makebox[1 mm]{}=\makebox[1 mm]{}
\pmatrix{& 0             & -i \sigma_{_2}  \cr 
         & -i \sigma_{_2}   &  0           \cr} $ ;   
the minus sign on RHS can be interpreted as the phase factor came from 
{\it Pauli exclusion\/} principle as in Sec.(1.1).  With a normalization 
condition given for massless particles(neutrinos)\cite{Itzykson-2}, that is,  
$ \nu_{_R}(k)^{\dag} \nu_{_R}(k) =  \nu_{_L}(k)^{\dag} \nu_{_L}(k) =  
 \bar \nu_{_R}(k)^{\dag} \bar \nu_{_R}(k) =  \bar \nu_{_L}(k)^{\dag} \bar 
\nu_{_L}(k)  = 2 E_{_k}$,  plane wave solutions in Standard representation 
can be constructed as 
$$
\nu_{_R}(k) = \sqrt{E_{_k}} \left(\begin{array}{c}
           +1 \\ 0 \\ +1 \\ 0 \end{array}\right) 
\makebox[2cm]{} 
\bar \nu_{_R}(k) = \sqrt{E_{_k}} \left(\begin{array}{c}
           -1 \\ 0 \\ -1 \\ 0 \end{array}\right) 
$$ 
$$
\nu_{_L}(k) = \sqrt{E_{_k}} \left(\begin{array}{c}
           0 \\ +1 \\ 0 \\ -1 \end{array}\right) 
\makebox[2cm]{} 
\bar \nu_{_L}(k) = \sqrt{E_{_k}} \left(\begin{array}{c}
           0 \\ -1 \\ 0 \\ +1 \end{array}\right) 
$$ 
\section{Neutrino}
If neutrino mass is zero as assumed in the last Section, there are only two 
physically distinctive neutrinos for each kind, and their helicities are 
uniquely fixed to {\it right-handed\/} or {\it left-handed\/} once for all  
because of $m_{\nu} = 0$.  Moreover, if we assume that $\nu_{_R}$ and 
$\bar \nu_{_L}$  exist, only the helicity of neutrino should be considered to 
distinguish neutrinos for each kind; and Dirac equation is closed by parity 
operation like other spin ${1\over 2}\hbar$ fermions.  Phenomenological facts 
-- neutrinos have an extremely small cross section, and Electromagnetic 
Interaction is highly suppressed in Weak Interaction limit($\sim 10^{-18} m$) 
-- can be a good clue to understand neutrino.  In what follows, we are going 
to propose what is {\it neutrino\/} and what is a possible model of $W^{\pm}$ 
boson with which {\it parity violation\/} can be explained. 
\subsection{Propagation mode}  
Since Electromagnetic wave, on which rest mass cannot be defined, was 
considered as {\it vacuum string oscillation\/}\cite{Kim-1}, we might consider 
a similarity of neutrino to photon.  In the model of electromagnetic wave 
propagation\cite{Kim-1}  Plank's constant($h$) is given as 
\begin{equation}
h = 2 \pi^2 m_e c \left( {A^2 \over d} \right), 
\label{Plank} 
\end{equation}
where $A$ is the {\it constant\/} amplitude for each wave string, $d$ is the  
distance between two adjacent vacuum electrons, and $m_e$ is bare electron 
mass.  
For a rough estimation if we use electron mass instead of the bare mass,  
electron Compton wavelength($\lambda_c /2 \pi$) has  relation as following 
\begin{eqnarray} 
{S \over d} & \approx  & {\hbar \over {m_e c}},  \nonumber \\
            & = & 3.86 \times 10^{-13} m, 
\label{P_C} 
\end{eqnarray}  
in which $S = \pi A^2$.  Now, if we accept $A$ as $\sim 10^{-18} m$ 
\footnote{in facts, it can be even bigger than $\sim 10^{-18} m$.}  because 
Electromagnetic Interaction is highly suppressed in this limit and, thus, we 
can surmise that electromagnetic wave cannot propagate in the limit.  
Then, $d$ is estimated as $\sim 10^{-23} m$ from Eqn.(\ref{P_C}).   
Furthermore, there is a limit of photon energy because $d$ is not zero --  
the vacuum string oscillation is not through a continuous medium.   
For example, wavelengths of T$eV$  and P$eV$ energy photons\footnote{T$eV = 
10^{12} eV$, P$eV = 10^{15} eV$, E$eV = 10^{18} eV$} are $\sim 10^{-18} m$ and 
$\sim 10^{-21} m$, respectively.  But E$eV$ energy photon is not possible 
because the wavelength($\lambda \sim 10^{-24} m $) is smaller than $d$.  
As a reference, P$eV$ energy order $\gamma$ rays have been detected in 
cosmic ray experiments.\cite{PEV}  
\par   
For neutrino to propagate Weak Interaction region($\sim 10^{-18} m$) there is 
a mode in which only two independent states exist, that is {\it longitudinal 
vacuum string oscillation\/} in which spin state(helicity) is also transferred 
through the string oscillation.  With this model we can estimate bare electron 
size by a crude comparison of Compton scattering with neutrino and electron 
elastic scattering($\nu_e, e$).  Let us say, the ratio of Compton scattering 
total cross section\footnote{low energy $\gamma$ ray to avoid hadronic 
interaction} to the elastic scattering cross section $\sigma({\nu_e}  e)$ is  
$$ 
R = {\sigma_{_{com}} \over \sigma_{_{\nu e}}} \sim {A^2 \over \delta^2}, 
$$ 
where $\delta$ is bare electron radius.  For 1 M$eV$ of $\nu_e$ and photon  
in electron rest frame, $\sigma_{_{com}} \sim 10^{-29} m^2$ and 
$\sigma_{_{\nu e}} \sim 10^{-48} m^2$.\cite{John}  Then, the ratio, 
$R \sim 10^{19}$, and the bare electron radius  can be estimated as $\delta 
\sim 10^{-27} m$.  
\subsection{$W^{\pm}$ boson and parity violation} 
\begin{figure}[h] 
\begin{center}
\leavevmode
\hbox{%
\epsfxsize=3.6in
\epsffile{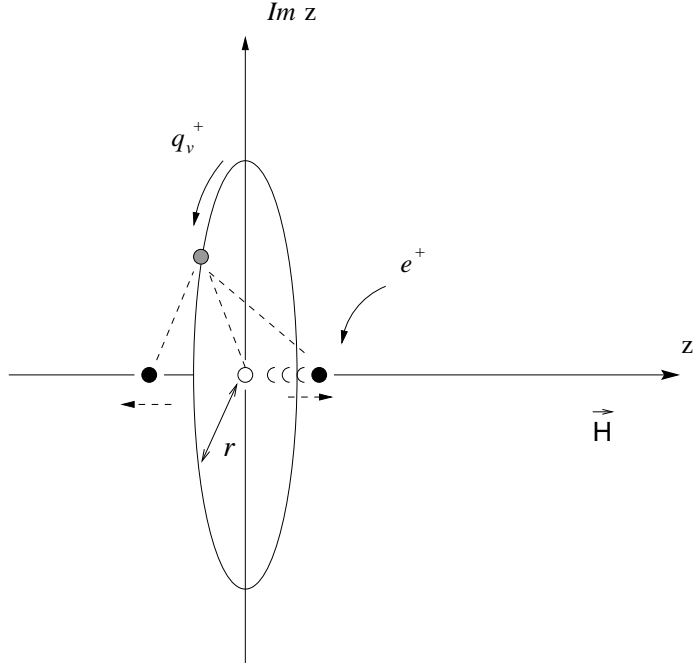}} 
\end{center}
\caption{$W^+$ in $\beta^+ $ decay} 
\label{W_boson}
\end{figure}

Through experimental facts we have known that $\bar \nu_{_L}$ and $\nu_{_R}$  
don't  exist in nature. Yet we need to distinguish the existence in 
phenomenology and in ontology.  If $\bar \nu_{_L}$ and $\nu_{_R}$ exist,  
necessarily we need an explanation why these two neutrinos couldn't have been 
detected. For example, in $\beta^+$ decay of ${^{8}B} \rightarrow  
{^{8}Be^*} + e^+  + \nu_e$ one proton is converted to a neutron inside 
the nucleus with emitting a positron $e^+$ and an electron neutrino $\nu_e$.  
That is $p \rightarrow n + e^+ + \nu_e$. In this process of Weak Interaction  
it has been known that $W^+$ boson intermediate the process and the neutrino 
$\nu_e$ is {\it left handed\/} electron neutrino.  
However, if we assume $W^+$ boson($\Gamma^{e\nu}_{_W} \simeq 0.23$ G$eV$, 
$\tau \sim 3 \times 10^{-24} sec$\cite{Gordon})  as a momentary interacting 
state of a positron and {\it virtual positive charge strings\/} moving around the positron,  we can 
figure out from Fig.(\ref{W_boson}) that there is no more preference in the 
neutrino helicity if the magnetic field is turned off. 
\paragraph{}
In Fig.(\ref{W_boson})  $e^+$ represents bare charge of the positron, and 
$q^+_v$ stands for the virtual positive charge strings induced from vacuum 
polarization\cite{Sakura}.  In which the virtual positive charge strings might 
even have a distribution depending on radial distance from the positron.  
Although we don't expect a positively induced  virtual charges from the vacuum 
polarization if the positron is in a free space, inside Weak Interaction 
region($r \leq 10^{-18} m$) the virtual positive charges can be accumulated 
around the edge of the region and even experienced a repulsive force from the 
edge.  According to the mirror image in Dirac's hole theory  
the virtual-positive-charges behave like that they have {\it positive\/} 
masses in phenomena.   
Since the state, interacting between the positron 
and virtual-positive-charges, is not stable, soon it decay to a free 
positron and a neutrino, in which the kinetic energy of 
the virtual-positive-charges is transferred to the longitudinal vacuum string 
oscillation, that can be interpreted as neutrino as assumed before.  
Now, if magnetic field $\vec H$ is applied as in Fig.(\ref{W_boson}) to investigate 
the positron's spin orientation  we can use Faraday Induction law, which is 
an empirical and macroscopic law, because the radius of $10^{-18} m$ is much 
bigger than the bare electron size($10^{-27} m$).  
Then, the virtual-positive-charges rotating around the positron  move to 
the negative z-axis(attractive in Faraday Induction law) and finally 
it will be transferred to the longitudinal vacuum string oscillation -- 
neutrino -- following negative imaginary z-axis 
as in the case of photon\cite{Kim-1} because neutrino mass $m_{_{\nu}}$ is 
assumed to be zero.  
On the other hand, the positron must move to the positive z-axis 
because of momentum conservation. If total spin of the system was $1  \hbar$ 
pointing to the positive z-axis, the emitting electron neutrino definitely has 
{\it left handed\/} helicity.   What if the magnetic fields is turned off? 
we cannot say which direction the neutrino choose for its emitting.  
From a reasonable guess, we can say that it should be equally probable 
for each direction.   
\par
\medskip
In $\beta^-$ decay, that is $n \rightarrow p + e^- + \nu_e$, $W^-$ boson also 
can be treated similarly as in $W^+$ boson case if we remind of the mirror 
image in Dirac's hole theory, where $W^-$ boson is considered as a momentary  
state of an electron $e^-$ and virtual-negative-charge strings moving around 
the electron, and the virtual-negative-charge strings behave like that they 
have negative masses.  With a similar set up like Fig.(\ref{W_boson}) except 
negative charges and the other rotating direction of the virtual-negative-
charge strings,  we can find out that the emitting electron neutrino is now 
{\it right handed\/} if the magnetic fields is on, and equally probable to be 
{\it right handed\/} or {\it left handed\/} if the magnetic field is off. 
Similarly, we can assume further that $Z^o$ boson is a momentary interacting 
state($\sim 3 \times 10^{-24} sec$) between virtual-positive-charge strings and 
virtual-negative-charge strings inside the Weak Interaction 
region($r \le 10^{-18} m$).       
\paragraph{}
From the reasoning as above, it is reasonable to say that the {\it parity 
violation\/} can be happened  if we consider only one part of phenomena, 
in which we can investigate the spin orientation of lepton using magnetic 
field; however, {\it intrinsically\/} parity is conserved 
if vacuum is magnetic field free.  
\subsection{neutrino flavors and solar neutrinos}
From experimental results\cite{Dandy} we can be sure that there are 
at least three kinds of flavors -- $\nu_e, \nu_{\mu}, \nu_{\tau}$. Up to now, 
it seems that we have assumed only one kind of neutrino because we cannot 
distinguish neutrino flavors with one longitudinal vacuum string oscillation 
model.   Yet, there is a way to distinguish them.  If neutrino propagation 
is assumed as a bundle of vacuum string oscillations in general, each flavor 
of neutrinos can be distinguished as a bulk motion with a different number of 
vacuum string oscillations. 
\paragraph{}
If there exist the counter parts of left-handed neutrino  $\nu_{_L}$ and 
right-handed antineutrino $\bar \nu_{_R}$ as assumed before, the parity should 
be conserved.    Moreover, we found in the model of  $W^{\pm}$ bosons 
that the helicity of neutrino is affected by the external magnetic field. 
\par
\medskip 
With these facts, let us try to find a possible explanation for the solar 
neutrino problem as mentioned before.  For example, $^8$B-neutrino 
deficit\cite{Sun} as 
${\Phi_{_{exp.}} \over \Phi_{_{SSM}}} \approx 0.47 \pm 0.10(1 \sigma)$ 
was confirmed again by Super-Kamiokande\cite{Kamio} experimental result. 
If solar magnetic field doesn't affect on helicities of neutrinos,  
the half of the neutrinos are {\it right handed\/};  the other half,  
{\it left handed\/}. Now, if we compare the cross-sections(leptonic 
interaction) of {\it right handed\/} neutrino and {\it left handed\/} neutrino 
to electron, 
the ratio of $\sigma(\bar\nu_{_R} e^-)$ to $\sigma(\nu_{_L} e^-)$ is     
$\sim 0.416$ for $^8$B-neutrinos.\cite{Winter}  Hence, with this fact we can 
estimate that $\Phi_{_{exp.}}$ should be 
$\sim 3.65 \times 10^6 cm^{-2} s^{-1}$.  In fact, it is still bigger than 
the experimental result of $2.42\pm0.06 \times 10^6 cm^{-2} s^{-1}$.  
\par
\medskip 
Although it is not easy to estimate how much solar magnetic field affect on 
solar neutrinos(helicity and  radiational direction), we can suppose that 
the neutrino flux itself from the Sun is not {\it isotropic\/}. 
In the model of $W^+$ boson in Fig.(\ref{W_boson}), we considered that virtual 
positive-charge-string is affected during the disintegration  to $e^+$ and 
$\nu_e$ ($\sim 10^{-24} sec$) -- Faraday Induction law.  If neutrinos, those of which we 
are looking for on earth, emit from the equatorial zone of the Sun,   
then  the direction of emission possibly can be deviated from 
the equatorial plane of the Sun.  This effect should reduce the neutrino flux 
to $66 \%$.     
Moreover, we can expect that the smaller energy neutrinos, the more strong deviations. 
In Kamiokande experimental result(1994)\cite{Kamio}, we can confirm this effect by 
comparing neutrino spectrums in the measurements and from the Monte Carlo 
simulation, that is the expectation from standard solar model.               
\section{Summary}
\paragraph{}
It has been known that Dirac equation represents spin ${1\over 2} \hbar$ 
fermions. Through this letter we compared Dirac formalism and Majorana 
formalism for neutrino case.  Before that, we investigated how Pauli exclusion 
principle is related to fermions in the Complex Space\cite{Kim-1} and 
if neutrino oscillation is physically feasible or not.  In short, neutrino 
oscillation is not compatible with Special Theory of Relativity.  
That is, as long as Special Theory of Relativity is impeccable, neutrino 
flavor oscillation is not possible. 
\par
\medskip 
\noindent                 
For neutrino, Majorana type neutrino was rejected in considering experimental 
results\cite{Cowan}\cite{LA}. If neutrino has rest mass, it should be treated 
like other spin ${1\over2} \hbar$ fermions.  However, we assumed that neutrino 
has no mass; right-handed neutrino $\nu_{_R}$ and left-handed 
antineutrino $\bar \nu_{_L}$ exist. 
Here, we use neutrino and antineutrino, but they are not different even though 
we use them traditionally in Dirac four spinor formalism.  There are only two 
physically distinctive neutrinos for each kind -- {\it right-handed \/} and  
{\it left-handed \/}.  From this fact and under the assumption that neutrino has 
no rest mass, the longitudinal vacuum string oscillation was suggested for 
neutrino.   Moreover, the intermediate bosons $W^{\pm}$ was also suggested 
as a model because it is necessary that {\it parity\/} should be conserved  
if right-handed neutrino $\nu_{_R}$ and left-handed antineutrino 
$\bar \nu_{_L}$ exist.  In the model of $W^{\pm}$ bosons,  we also found how 
we might have overlooked the truth.   
Meanwhile, V-A theory, with which the interaction Hamiltonian in Weak Interaction 
has been formulated, is not enough to include $\nu_{_R}$ and  $\bar\nu_{_L}$. 
Instead,  both V-A and V+A theories should be considered 
because we are assuming {\it parity\/} conservation -- 
that means existing the mirror image.   
Moreover, the {\it lepton number\/} conservation law should be extended 
to include the other part, that might be a pair of {\it lepton number\/} 
conservations, separately.               
\paragraph{} 
The neutrino oscillation has been considered as a candidate to a new physics 
to explain the discrepancy of standard solar model\cite{Sun} in  
solar neutrino experiments.\cite{William}\cite{SNE}     
In spite of that, we investigated {\it neutrino\/} itself alternatively 
and suggested one way to solve the solar neutrino problem.  

\end{document}